**Parkinson's disease is a TH17 dominant autoimmune disorder against accumulated alpha-synuclein**

By Wan-Chung Hu [*]

*Postdoctorate
Genomic Research Center
Academia Sinica
128 Academia Road section 2
Nangang 115, Taipei
Taiwan

Current institute:

*Department of Neurology
Shin Kong Memorial Hospital (Medical Center)
Taipei, Taiwan

**Abstract**

Parkinson's disease is a very common neurodegenerative disorder. Patients usually undergo destruction of substantia nigra to develop typical symptoms such as resting tremor, hypokinesia, and rigidity. However, the exact mechanism of Parkinson's disease is still unknown, so it is called idiopathic Parkinsonism. According to my microarray analysis of peripheral blood leukocytes and substantia nigra brain tissue, I propose that Parkinson's disease is actually a TH17 dominant autoimmune disease. Based on the microarray data in substantia nigra, HSP40, HSP70, HSP90, HSP27, HSP105, TLR5, TLR7, CEBPB, CEBPG, FOS, and caspase1 are significantly up-regulated. In peripheral leukocytes, NFKB1A, CEBPD, FOS, retinoic receptor alpha, suppressor of IKK epsilon, S100A11, G-CSF, MMP9, IL-1 receptor, IL-8 receptor, TNF receptor, caspase8, c1q receptor, cathepsin Z, HLA-G, complement receptor1, and complement 5a receptor. General immune related genes are also up-regulated including ILF2, CD22, CD3E, BLNK, ILF3, TCR alpha, TCR zeta, TCR delta, LAT, ITK, Ly9, and BANK1. The autoantigen is mainly alpha-synuclein. After knowing the exact disease pathophysiology, we can develop better drugs to prevent or control the detrimental disorder.

**Introduction**

Parkinson's disease is the second most common neurodegenerative disorder. Primary Parkisonism is also called idiopathic Parkinson's disease because its exact etiology is still unknown. After the destruction of substantia nigra of midbrain, patients will develop resting tremor, cogwheel rigidity, hypokinesia, and postural instability. Patients' life expectancy is much shorter than normal healthy population. Even it is a very common and detrimental illness, the etiology and pathogenesis is still a puzzle. By using microarray analysis of peripheral blood leukocytes and substantia nigra brain tissue, I find out that idiopathic Parkinsonism is actually a TH17 dominant autoimmune disease. Summary of all previous literatures also support that Parkinson's disease is a TH17 autoimmune illness.

**Materials and methods**

Microarray datasets of Parkinson's disease

In this paper, I incooperate two microarray datasets for analysis. The fist one is peripheral blood mononuclear cell gene expression profiles in Parkinson's disease patient. The dataset GSE6613 is available in GEO website. Dr. C.R. Scherzer published the research results of peripheral blood biomarkers for Parkinson's disease in PNAS. The second dataset(GSE7621) is substantia nigra gene expression profiles in Parkinson's disease. Substantia nigra is the major affected brain area of Parkinson's disease.

Statistical analysis

RMA express software(UC Berkeley, Board Institute) is used to do normalization and to rule out the outliners of the above dataset. I rule out the potential outliners of samples due to the following criteria:
1. Remove samples which have strong deviation in NUSE plot
2. Remove samples which have broad spectrum in RLE value plot
3. Remove samples which have strong deviation in RLE-NUSE mutiplot
4. Remove samples which exceed 99% line in RLE-NUSE T2 plot

After the analysis of RMA express, we select 47 Parkinson's disease and 19 healthy controls. We remove GSE153479, GSE153497, and GSE153508 in healthy control due to RNA express analysis. We remove GSE153411, GSE153421, GSE153454 in Parkinsonism group. (Figure 1 and Figure 2) Then, we use Genespring to find out statistically significant expressed genes in substantia nigra and peripheral leukocytes of Parkinson's disease.

Validation by RT-PCR

Since this study is a secondary dataset analysis, I don't have samples in hand to do wet bench validation such as RT-PCR. However, in Dr. Scherzer's paper, he performed RT-PCR to confirm his result of microarray experiment by using RT-PCR. He found that the gene ST13, a HSP70 cofactor, is down-regulated in Parkinsonism compared to that in Healthy control due to microarray analysis. Then, he used quantitative real-time PCR assays based on precise fluorogenic 5' nuclease chemistry in a large set of 39 Parkinsonism patients and 12 age-, sex-, and blood count- matched healthy controls. He used the housekeeping gene glyceraldehydes-3-phosphate dehydrogenase to control for input RNA and the comparative threshold cycle method for analysis. The RT-PCR results confirmed that ST13 is lower in Parkinsonism patients. Thus, it states that the microarray platform of Dr. Scherzer's experiments is valid and reliable.

**Results**

Heat shock protein, Toll-like receptor, Caspases, and TH17 related gene up-regulation in substantia nigra of Parkinson's disease patients

In the first dataset, we selected 816 genes of substantia nigra of Parkinson's disease compared to healthy control with fold change>1.5 and Benjamin adjusted false discovery rate<0.05. Then, we selected immune-related genes from this 816 genelist.(Table1) We find out that heat shock protein genes are almost all up-regulated in Parkinson's disease including HSP40, HSP70, HSP90, HSP27, and HSP105. In addition, TLR5 and TLR7 are up-regulated as well. Based on several references, TLR5 and TLR7 can recognize heat shock protein to trigger TH17 immunity. Several toll-like signaling molecules are also up-regulated in Parkinson's disease. In addition, many TH17 related transcription factors are up-regulated in Parkinson's disease including CEBPbeta, CEBPgamma, and FOS(AP1). STAT1, the key transcription factor for TH1 and TH9 immunity, is downregulated in Parkinson's disease patients. Key TH17 cytokine receptors are also up-regulated including IL-17 receptor and TGF beta receptor. And, TH2 cytokine receptor-IL13 receptor is down-regulated in Parkinson's disease substantia nigra. Finally, the key downstream effector molecule caspase1 of TH17 cytokine IL-1 is up-regulated in Parkinson's patients. This caspase 1 up-regulation may closely related to the neuron death triggered by TH17 cytokine IL-1.

TH17 related molecules are up-regulated in peripheral leukocytes of Parkinson's disease patients

In our second analysis, we find out that 505 significantly expressed genes in peripheral leukocytes of Parkinson's patients with fold change>1.1 and unadjusted false discovery rate P<0.05. Then, we pick 103 immune related genes out of the 505 genelist. (Table2) We find out that many Th17 related molecules are up-regulated in WBC of Parkinson's disease patients.

First of all, many TH17 driven transcription factors are up-regulated in leukocytes of Parkinsonism patients. NFKB1A, CEBPD, FOS(AP-1), retinoic receptor alpha, and suppressor of IKK epsilon are up-regulated. They are master transcription factors for TH17 immunity. Retinoic receptor alpha is related to the differentiation of main TH17 effector cells-Neutrophils. Besides, the TH9 (THαβ) related transcription factor IRF8 is down-regulated.

Second, many TH17 effector molecules are also up-regulated in peripheral leukocytes of Parkinson's disease. S100A11, LPS induced TNF factor, TNF alpha induced protein2, CD32, G-CSF, lysosomal mannosidase, MMP9, CD16b, type II interleukin 1 receptor, interleukin 8 receptor alpha & beta, TNF receptor 9 & 1A & 10C, TNF factor 14, casapase8, S100A11P, CD93 (c1q receptor), cathepsin Z, HLA-G, leukocyte immunoglobulin-like receptors, type1 thrombospondin containing 7A, complement receptor1, IgG Fc receptor transporter alpha, and complement 5a receptor are all up-regulated after Parkinson's disease. These up-regulated molecules highly suggest that TH17 immunological pathway is triggered in Parkinsonism. In addition, Fas inhibitory gene and BCL2 are down-regulated that suggest apoptotic signals are up-regulated in Parkinson's disease.

Lymphocyte related genes and immune genes other than TH17 pathway are down-regulated in leukocytes of Parkinson's disease. Generally, neutrophils are up-regulated and lymphocytes are down-regulated in TH17 immunity. Thus, this result can also help to point out Parkinson's disease is a TH17 dominant autoimmune disease. These lymphocyte related genes include ILF2, CD22, CD3E, BLNK, ILF3, TCR alpha, TCR zeta, LAT, ITK, LY9, BANK1, and TCR delta. TH1, TH9, and TH2 immune genes are down-regulated including GBP1, IRF8, CD44, KIR3DL2, HEBP1, HEBP2, and TRAF3. IL-32 which can suppress NFkB and STAT3(TH17 transcription factor) is also down-regulated in leukocytes of Parkinson's disease.

Unlike the up-regulation of heat shock proteins in substantia nigra, heat shock protein genes in peripheral leukocytes are down-regulated in Parkinson's disease patients. These heat shock protein related genes include HSP90AB1 and ST13. In addition, the main autoantigen of Parkinsonism-alpha synuclein gene is down-regulated in leukocytes of Parkinson's disease patients as well. There might be some negative feedback machinery for these down-regulated disease related genes.

**Discussion**

Idiopathic Parkinson's disease is a very common neurodegenerative disorder, but its etiology is still unknown. Although there are a few cases of familial Parkinson's diseases, sporadic Parkinson's diseases are still the major population of this disorder. Parkinson's disease is found to be associated with several risk factors such as MPTP exposure, heavy metal such as Manganese exposure, pesticides such as rotenone or paraquet exposure, herbicides such as Agent Orange exposure, and Influenza virus infection.(Betarbet, Sherer et al. 2000) Estrogen, anti-oxidants, caffeine, and NSAID have protective roles from Parkinson's disease.(Ross, Abbott et al. 2000; Chen, Zhang et al. 2003; Sawada and Shimohama 2003) Here, I propose that Parkinson's disease is actually a TH17 dominant autoimmune disease which can explain the above evidences.

Alpha synuclein is the major accumulated protein forming inclusion bodies in substantia nigra neurons of Parkinson's patients. In familial Parkinsonism, autoantibody against mutation of alpha synuclein can contribute to the disease pathogenesis.(Papachroni, Ninkina et al. 2007) Alpha synuclein is a kind of acute response amyloid protein which can be up-regulated by Th17 cytokine IL-1 during some bacterial infection.(Griffin, Liu et al. 2006) Similar to other neurodegenerative diseases such as Alzheimer's disease, the accumulation of alpha synuclein can up-regulate molecular chaperons such as heat shock protein 60 or heat shock protein 70 to trigger toll-like receptor driven TH17 immunity. Heat shock proteins released by neuron or glia cells can activate TLR2/4/5/7 of immune cells to trigger TH17 cytokines or chemokines.(Asea, Kraeft et al. 2000) Neuron or glia cells can also posse Toll-like receptor to trigger TH17 cytokine such as IL-6 or TNF alpha release in neuron or glia cells.(Alvarez-Erviti, Couch et al. 2011) These cytokines can be autocrine to enhance more Th17 cytokine release. More immune effector cells will be recruited and release more TH17 cytokines and chemokines. TH17 cytokine IL-1 can in turns up-regulate more alpha synuclein in neuron or glia. That is a vicious cycle. In addition, TH17 cytokines such as IL-1, TNFα, and IL-6 can act on neuron cells to undergo Wallerian-like degeneration.(Saha and Pahan 2003) That is the pathophysiology of Parkinson's disease. This pathogenesis is very similar to other neurodegenerative disorders such as Alzheimer's disease and Huntington's disease.

Witebsky's postulates are the standard criteria to decide if a disorder is an autoimmune disease. There are three requirements in Witebsky's postulates: 1. Direct evidence from transfer of pathogenic antibody or T cells 2. Indirect evidence based on reproduction of the autoimmune disease in experimental animal 3. Circumstantial evidence from clinical clues. First, in an Acta Neurology paper, injection of IgG antibody from Parkinsonism patients can cause substantia nigra destruction and Parkinson-like disease in rats.(Chen, Le et al. 1998) In a Brain paper, all Parkinsonism patients have IgG autoantibody recognizing dopaminergic neurons.(Orr, Rowe et al. 2005) In addition, Fc gamma receptors are up-regulated in lymphocytes and microglia of Parkinson disease patients. In an Experimental Neurology paper, injection of autoreactive IgG antibody isolated from Parkinsonism patients can also cause substantia nigra destruction in mice.(He, Le et al. 2002) In a paper from Journal of Neuroinflammation, serum antibodies from Parkinsonism patients can react with neuron membrane of mice dopaminergic cell line.(Huber, Mondal et al. 2006) In the presence of microglia, the serum antibodies can also suppress the production of dopamine from the mice dopaminergic cell line. In a JCI paper, CD4 T cell infiltration is found in postmortem Parkinson disease brain tissue and is related to neurodegeneration.(Brochard, Combadiere et al. 2009) In another paper of Journal of Neuroinflammation, complement activation is noted at substantia nigra isolated from expired Parkinsonian patients compared to normal control.(Loeffler, Camp et al. 2006) In a review paper, elevated autoantibodies and complements are found in serum and CSF of Parkinsonians. In addition, anti-alpha synuclein autoantibody is noted in familial Parkinsonism patients. Cytokine IL-1, TNF, and IL-6 are also increased in the serum, CSF, and substantia nigra of Parkinson disease patients. CD4 T cells are also increased in serum of Parkinson disease patients. In Parkinsonism's substantia nigra, up-regulation of TH17 effector molecule iNOS and cycloxygenase are found.(Gatto, Carreras et al. 1996) As for the second criteria, autoreactive CD4 T cell infiltration is found in brain of mice Parkinsonism model, and these CD4 T cells contribute to neurodegeneration. In rat Parkinsonism model, CD4(+)CD25(+) Treg cells have neuroprotective roles and TH17 cells are detrimental(Reynolds, Stone et al. 2010). In mice Parkinsonism model, overexpression of human alpha synuclein can lead to microglia activation and adaptive immune reaction.(Zhang, Wang et al. 2005) In a Brain paper, local and systemic IL-1 can exacerbate neurodegeneration and motor symptoms in animal models of Parkinsonism.(Pott Godoy, Tarelli et al. 2008) In a Brain Research paper, IL-1 elevation induced by MPTP cause neurodegeneration in MPTP Parkinsonism mice models.(Bian, Li et al. 2009) In a Glia paper, TH17 effector cells-neutrophil infiltration in substantia nigra causes Parkinsonism-like symptoms. In addition, mice deficient in TNF receptor are protected from Parkinsonism symptoms. Systemic and local injection of extracellular bacteria product LPS can produce neurodegeneration including Parkisonism symptoms.(Meredith, Sonsalla et al. 2008) Key TH9 (THαβ) cytokine IL-10 has protective role in rat model of Parkinsonism.(Johnston, Su et al. 2008) It is worth noting that alpha synuclein knockout mice didn't develop Parkinsonism after MPTP injection.(Dauer, Kholodilov et al. 2002) MPTP can cause alpha synuclein up-regulation and aggregation in substantia nigra.(Vila, Vukosavic et al. 2000) Thus, alpha synclein is the major mediator of MPTP induced Parkinsonism. Autoimmune against alpha synuclein is the most likely pathogenesis. As for criteria 3, there is association between Parkinson's disease and HLA haplotypes.(Hamza, Zabetian et al. 2010; Nalls, Plagnol et al. 2011) Besides, polymorphisms of TH17 related cytokines such as IL-1α, IL-1β, TNF-α, TNF receptor, IL-1 receptor antagonist,

CD14 receptor, and IL-6 are related to the risk of Parkinson's disease.(Wahner, Sinsheimer et al. 2007) Elevated serum or CSF key TH17 cytokine IL-1 and IL-6 is also associated with the increasing risk of Parkinson's disease.(Blum-Degen, Muller et al. 1995) Autoimmune seborrheic dermatitis is associated with the risk of Parkinsonism. Anti-inflammatory caffeine, NASID, or estrogen intakes have protective role for the Parkinson's disease development.(Giraud, Caron et al. 2010) Fetal stem cell transplant is newly used for Parkinson's disease treatment. However, graft rejection is noted and there is only short-term beneficial effect not long-term therapeutic effect for the fetal stem cell transplant.(Spencer, Robbins et al. 1992; Bjorklund, Dunnett et al. 2003; Bachoud-Levi, Gaura et al. 2006; Krystkowiak, Gaura et al. 2007) It is because there are still autoreactive immune cells attacking alpha synuclein positive neuron cells. Thus, the new transplant will still be rejected by the abnormal host immune system. These evidences all point out that idiopathic Parkinson's disease is an autoimmune disorder.

Parkinson's disease is TH17 dominant autoimmune disorder against alpha synuclein. Current treatment strategies include dopamine supplement and fetal stem cell transplant. However, these therapeutic methods could not solve the underlying pathogenesis of Parkinson's disease and stop the disease progression. We should use anti-inflammatory agents such as NSAID to treat. In addition, we need to develop new strategy to induce tolerance for alpha synuclein to stop Parkinson's disease. I sincerely hope we can control this detrimental illness in the future.

**Author's information**

Wan-Chung Hu is a MD from College of Medicine of National Taiwan University and a PhD from vaccine science track of Department of International Health of Johns Hopkins University School of Public Health. He is a postdoctorate in Genomics Research Center of Academia Sinica, Taiwan. His previous work on immunology and functional genomics studies were published at *Infection and Immunity* 2006, 74(10):5561, *Viral Immunology* 2012, 25(4):277, and *Malaria Journal* 2013,12:392. He proposed THαβ immune response as the host immune response against viruses.

Figure legends

Figure 1. RMA express plot for selecting samples in normal healthy controls.
1-A NUSE boxplot for normal control
1-B RLE boxplot for normal control
1-C RLE-NUSE multiplot for normal control
1-D RLE-NUSE T2 plot for normal control

Figure 2. RMA express plot for selecting samples in Parkinson's disease patients.
2-A NUSE boxplot for Parkinson's disease patients
2-B RLE boxplot for Parkinson's disease patients
2-C RLE-NUSE multiplot for Parkinson's disease patients
2-D RLE-NUSE T2 plot for Parkinson's disease patients

Figure 1

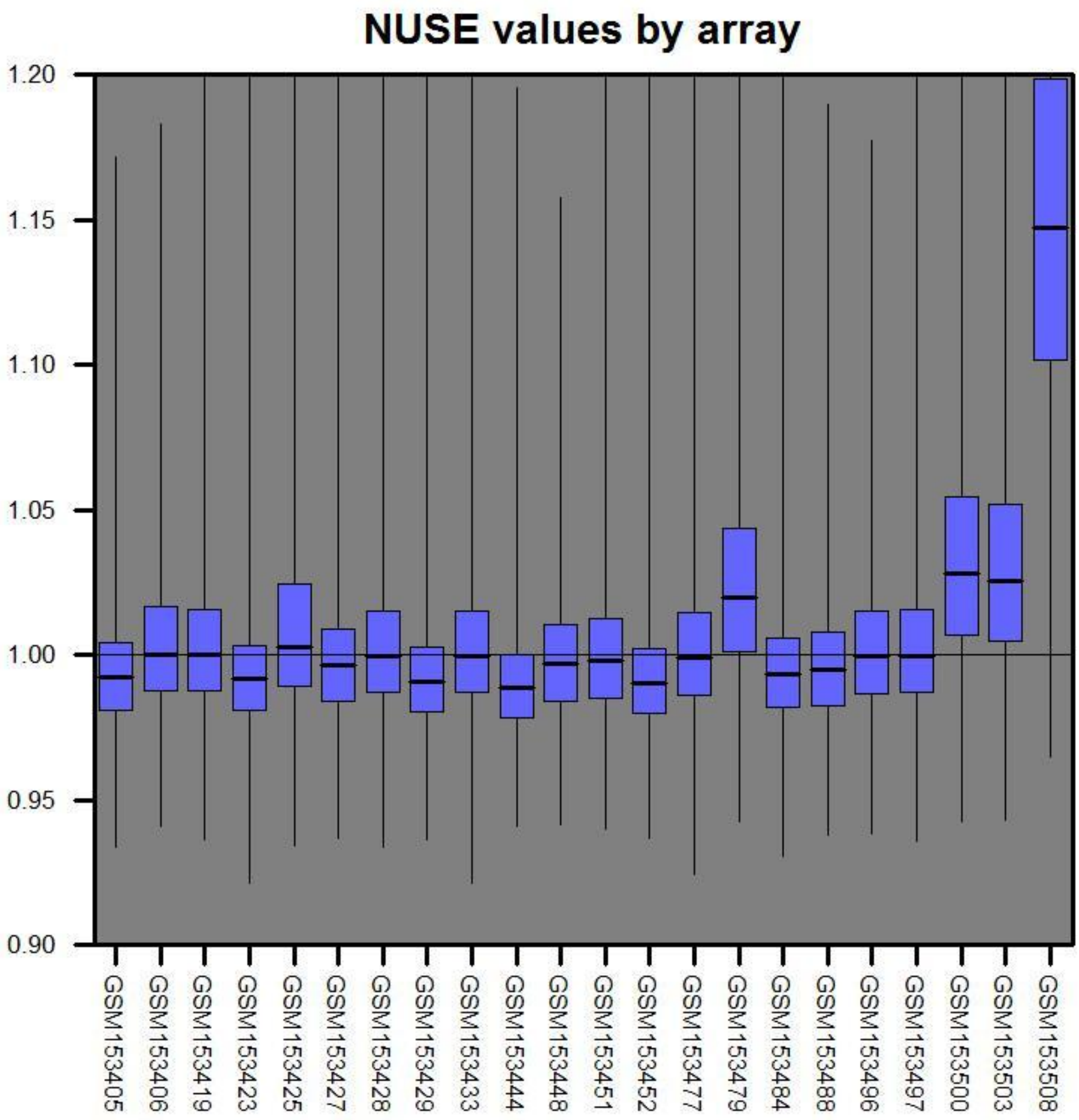

1-A

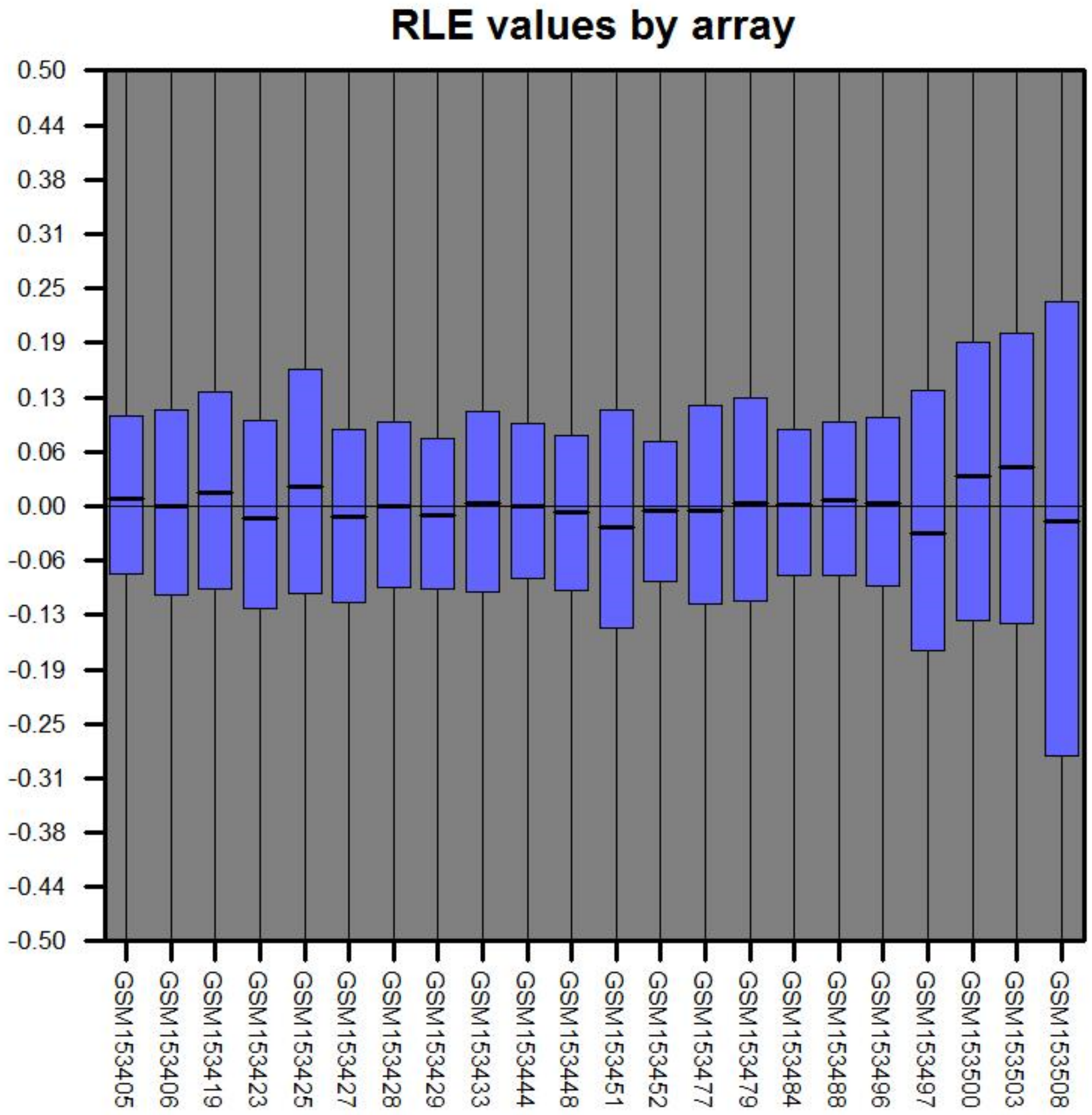

1-B

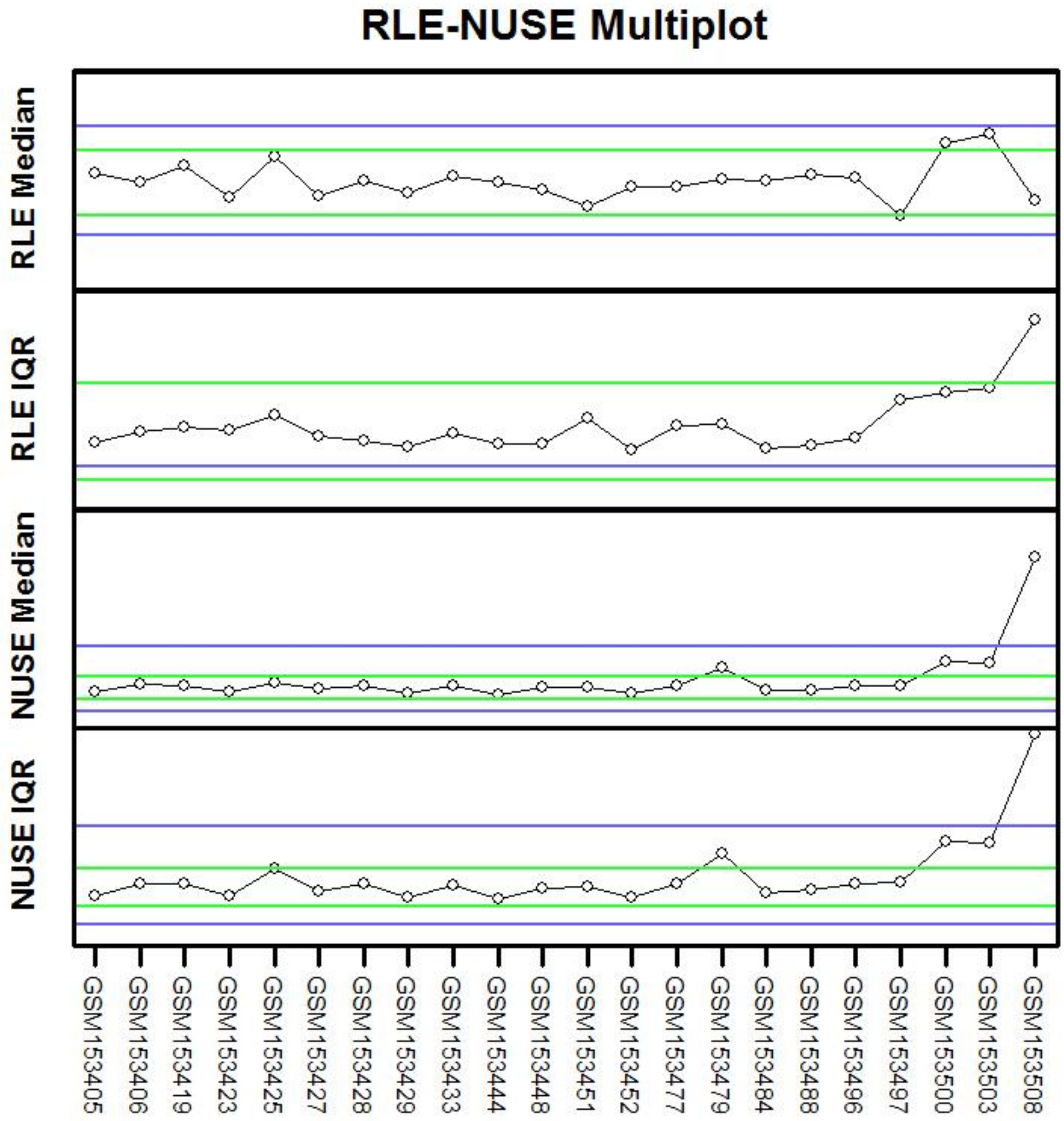

1-C

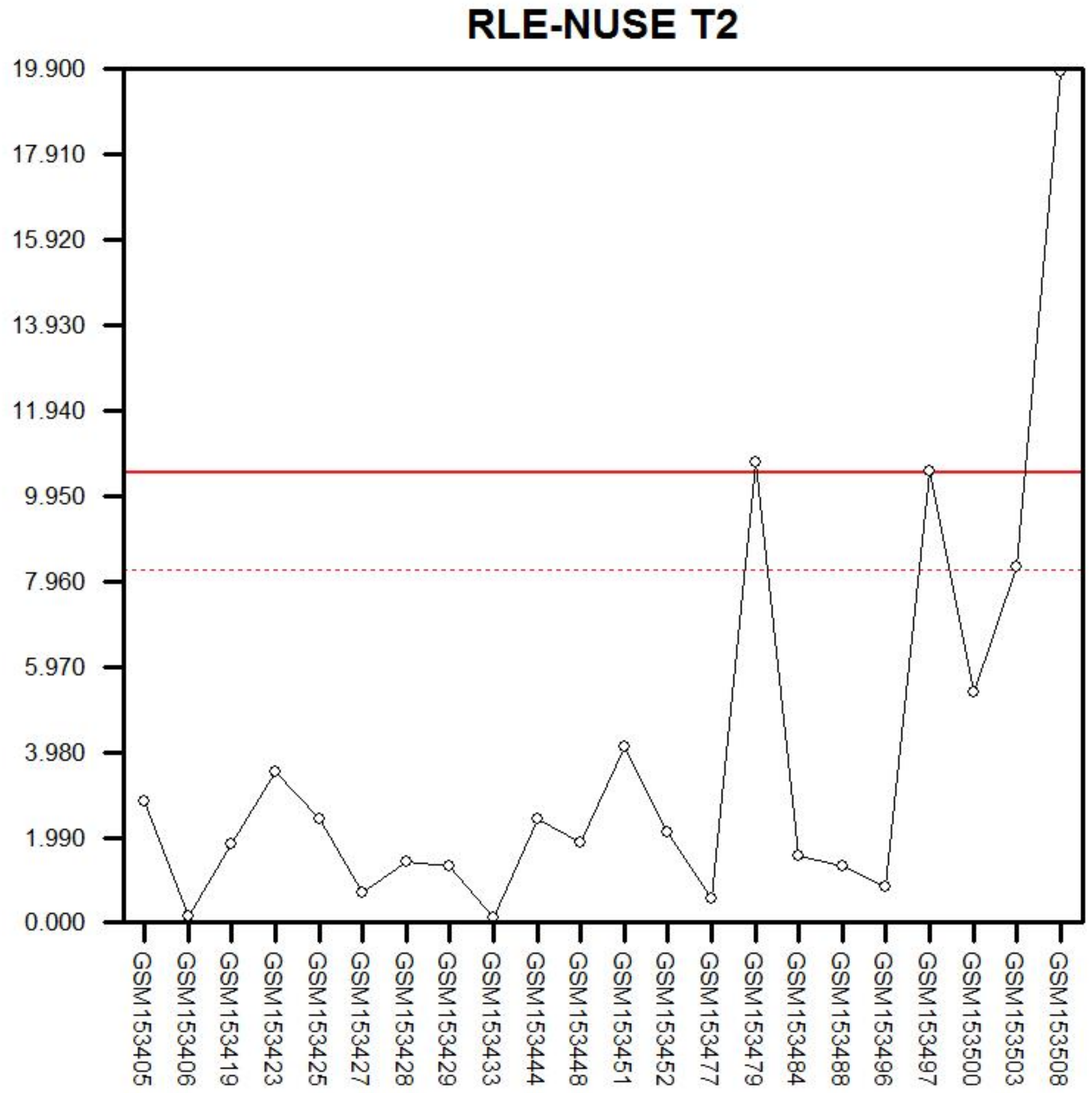

1-D

Figure 2

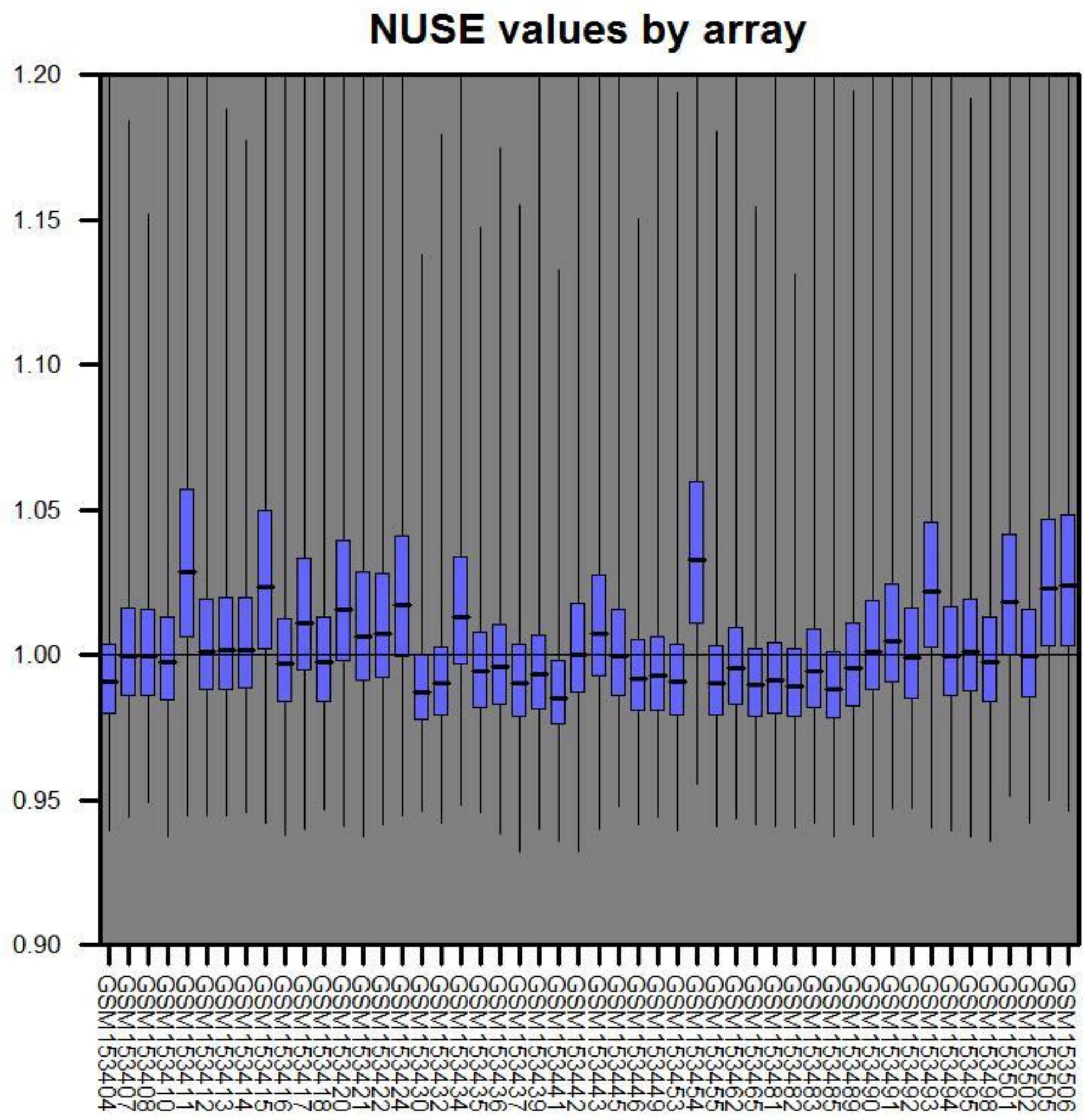

2-A

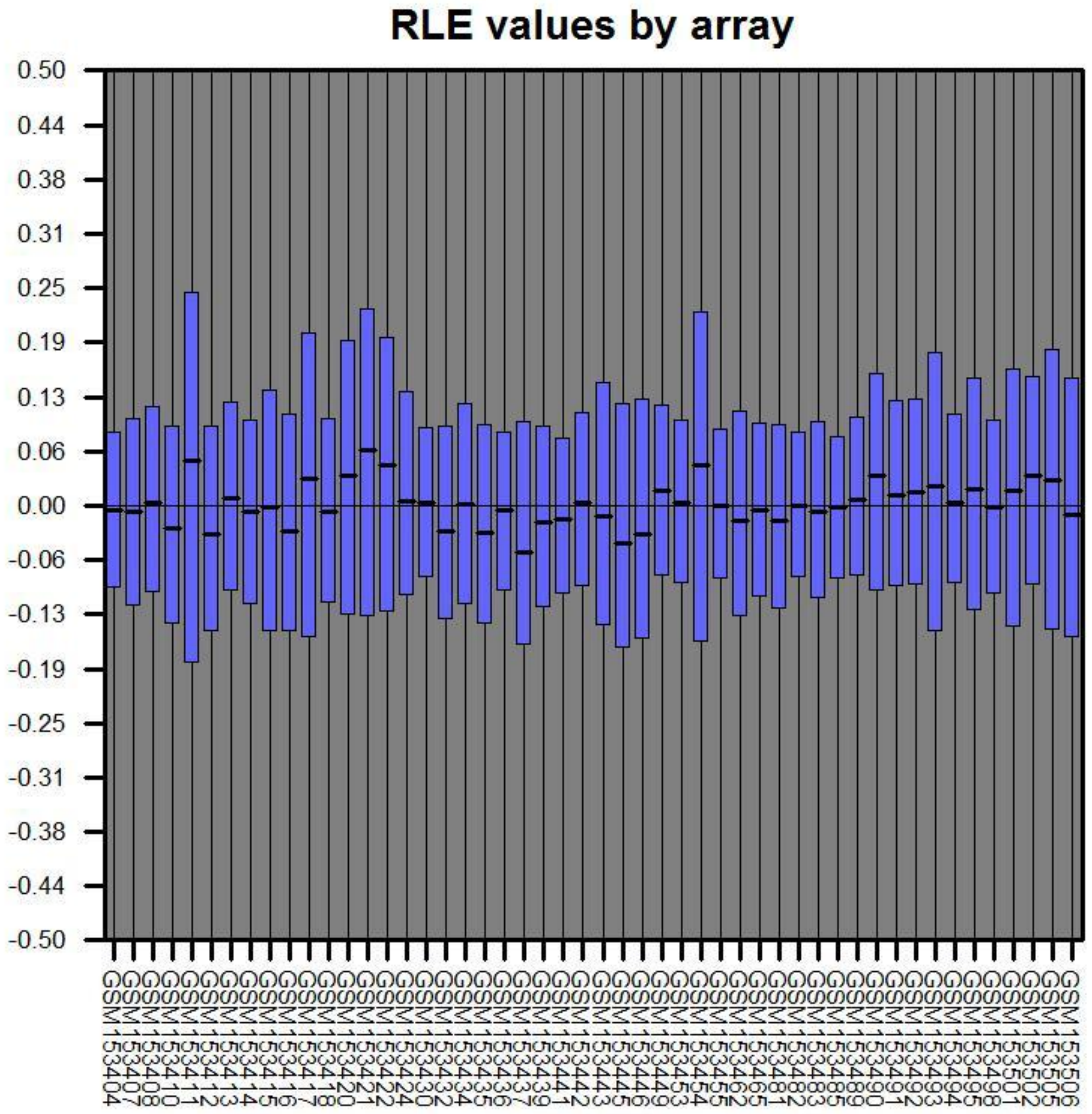

2-B

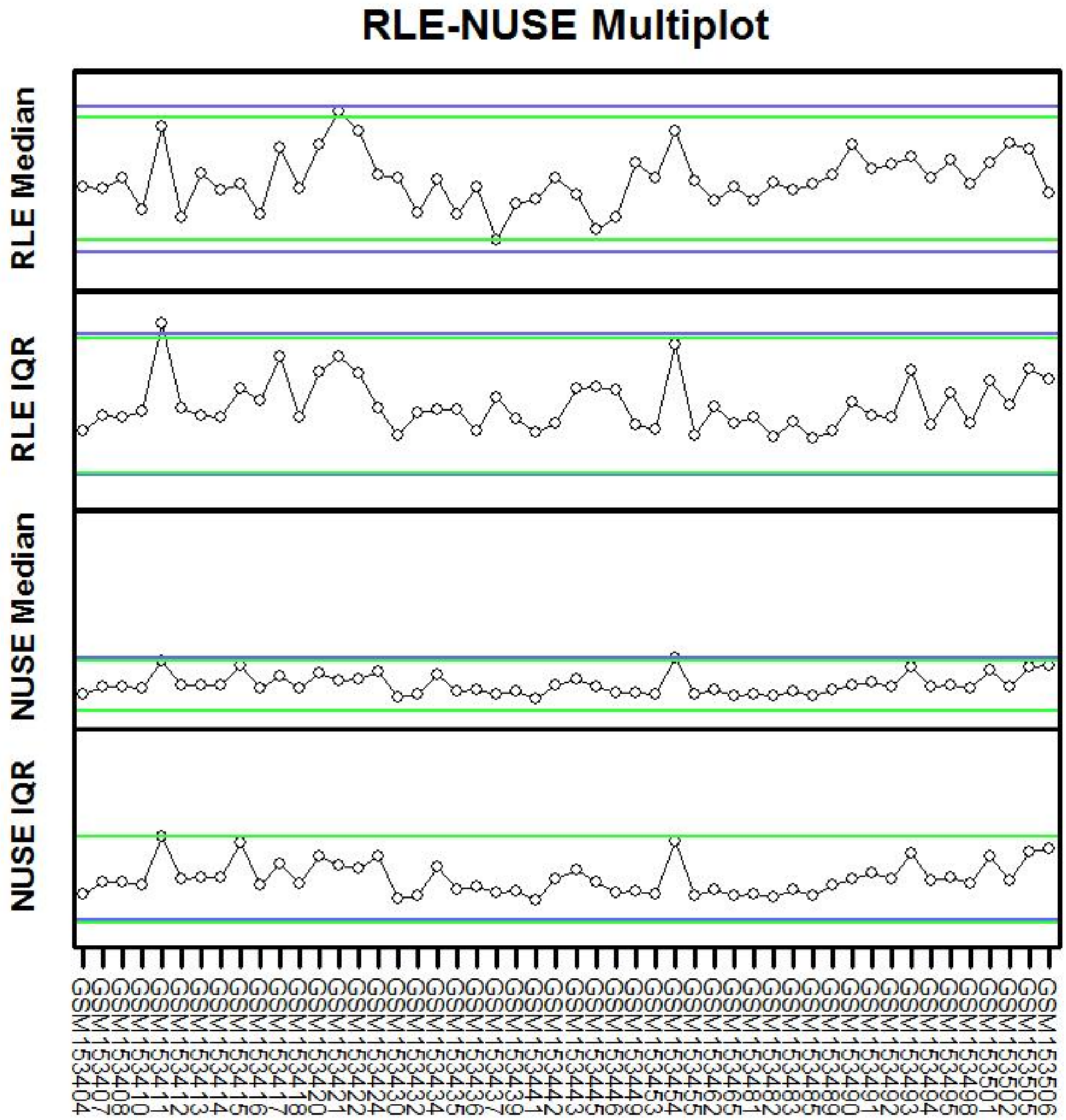

2-C

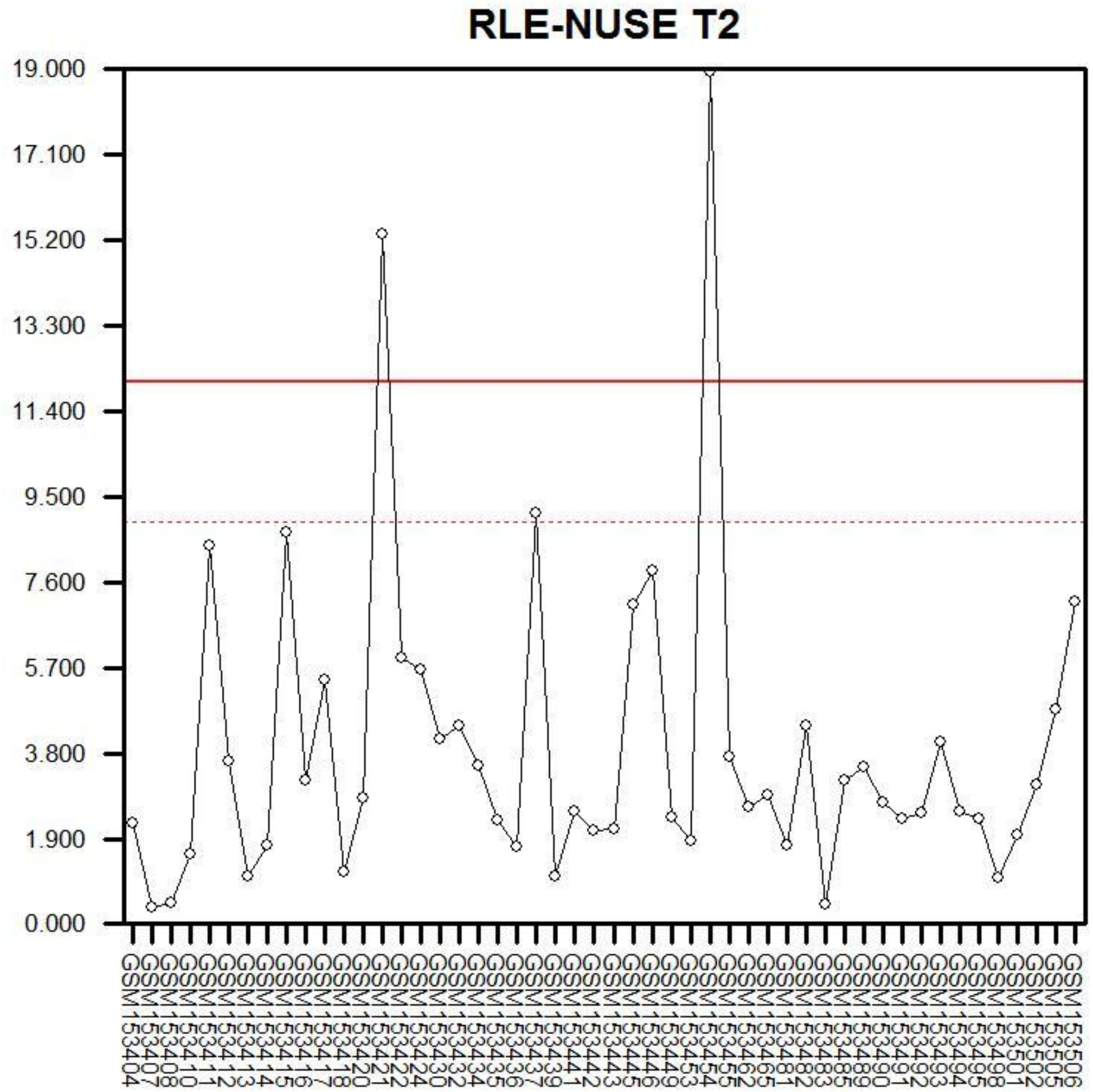

2-D

Table 1. Heat shock protein and TH17-related gene up-regulation in substantia nigra of Parkinson's disease patients

| Probeset | Pvalue | Arrow | Fold | Gene | Description |
|---|---|---|---|---|---|
| Heat Shock | | | | | |
| 200666_s_at | 0.046436 | up | 1.591286 | DNAJB1 | Hsp40B1 |
| 200799_at | 0.016182 | up | 1.942031 | HSPA1A | heat shock 70kDa protein 1A |
| 200800_s_at | 0.009898 | up | 2.568059 | HSPA1A/B | heat shock 70kDa protein 1A 1B |
| 200881_s_at | 0.025109 | up | 1.5808 | DNAJA1 | Hsp40A1 |
| 201491_at | 0.00584 | up | 1.565841 | AHSA1 | activator of heat shock 90kDa |
| 201841_s_at | 0.041056 | up | 1.93935 | HSPB1 | heat shock 27kDa protein 1 |
| 202581_at | 0.008943 | up | 2.568776 | HSPA1A/B | heat shock 70kDa protein1A /1B |
| 206976_s_at | 0.015495 | up | 1.938839 | HSPH1 | heat shock 105kDa/110kDa 1 |
| 209015_s_at | 0.001648 | up | 2.366288 | DNAJB6 | Hsp40B 6 |
| 210189_at | 0.001899 | up | 1.692562 | HSPA1L | heat shock 70kDa protein 1-like |
| 221782_at | 0.013571 | down | 1.539068 | DNAJC10 | Hsp40C10 |
| 227808_at | 0.01902 | down | 1.660796 | DNAJC15 | Hsp40C15 |
| 230148_at | 0.001627 | up | 1.813061 | AHSA2 | activator of heat shock 90kDa |
| Toll-like | | | | | |
| 210166_at | 0.014399 | up | 1.605882 | TLR5 | toll-like receptor 5 |
| 220146_at | 0.046749 | up | 1.506414 | TLR7 | toll-like receptor 7 |
| 228234_at | 0.002924 | up | 1.602183 | TICAM2 | toll-like receptor adaptor 2 |
| 231779_at | 0.014626 | up | 1.588289 | IRAK2 | IL-1 receptor-as kinase 2 |
| Transcription factor | | | | | |
| M97935MBat | 0.014721 | down | 1.52548 | STAT1 | signal transducer and activator1 |
| 212501_at | 0.047435 | up | 1.659838 | CEBPB | CCAAT/enhancer binding beta |
| 225527_at | 0.00591 | up | 1.503377 | CEBPG | CCAAT/enhancerbinding gamma |
| 228188_at | 0.012176 | up | 2.218706 | FOSL2 | FOS-like antigen 2 |
| 203574_at | 0.043302 | up | 1.816873 | NFIL3 | nuclear factor, interleukin 3 |
| Caspase | | | | | |
| 206011_at | 0.036084 | up | 1.51373 | CASP1 | caspase 1 |
| 209970_x_at | 0.040759 | up | 1.542726 | CASP1 | caspase 1 |
| Cytokine receptor | | | | | |
| 219255_x_at | 0.012304 | up | 1.755705 | IL17RB | interleukin 17 receptor B |
| 224156_x_at | 0.017114 | up | 1.627902 | IL17RB | interleukin 17 receptor B |
| 224361_s_at | 0.020254 | up | 1.631049 | IL17RB | interleukin 17 receptor B |
| 206172_at | 0.023187 | down | 1.60825 | IL13RA2 | interleukin 13 receptor, alpha 2 |
| 236561_at | 0.028704 | up | 1.605665 | TGFBR1 | TGF, beta receptor 1 |

Table 2. TH17 related gene up-regulation in peripheral leukocytes of Parkinson's disease patients

| Probe Set | Fold | Arrow | p-value | Gene |
|---|---|---|---|---|
| 200052_s_at | 1.188939 | down | 0.009691 | ILF2 |
| 200064_at | 1.141238 | down | 0.015928 | HSP90AB1 |
| 200660_at | 1.184608 | up | 0.025142 | S100A11 |
| 200704_at | 1.161244 | up | 0.010052 | LITAF |
| 200706_s_at | 1.168248 | up | 0.043356 | LITAF |
| 200986_at | 1.17499 | down | 0.02649 | SERPING1 |
| 201244_s_at | 1.164101 | up | 0.044455 | RAF1 |
| 201315_x_at | 1.132939 | up | 0.030281 | IFITM2 |
| 201502_s_at | 1.200774 | up | 0.017593 | NFKBIA |
| 202201_at | 1.200977 | down | 0.021401 | BLVRB |
| 202270_at | 1.28927 | down | 0.022318 | GBP1 |
| 202379_s_at | 1.223381 | up | 0.016804 | NKTR |
| 202388_at | 1.256398 | up | 0.00215 | RGS2 |
| 202509_s_at | 1.107859 | up | 0.013255 | TNFAIP2 |
| 202877_s_at | 1.125594 | up | 0.037677 | CD93 |
| 203041_s_at | 1.200253 | up | 0.01719 | LAMP2 |
| 203140_at | 1.211998 | up | 0.013618 | BCL6 |
| 203430_at | 1.122783 | up | 0.043456 | HEBP2 |
| 203561_at | 1.141576 | up | 0.033892 | FCGR2A |
| 203591_s_at | 1.21527 | up | 0.003722 | CSF3R |
| 203685_at | 1.214759 | down | 0.01472 | BCL2 |
| 203778_at | 1.104544 | up | 0.024774 | MANBA |
| 203828_s_at | 1.177826 | down | 0.036371 | IL32 |
| 203936_s_at | 1.261711 | up | 0.026599 | MMP9 |
| 203973_s_at | 1.265417 | up | 0.007709 | CEBPD |
| 204007_at | 1.119838 | up | 0.018176 | FCGR3B |
| 204057_at | 1.11098 | down | 0.03413 | IRF8 |
| 204466_s_at | 1.481427 | down | 0.022285 | SNCA |
| 204467_s_at | 1.244787 | down | 0.049072 | SNCA |
| 204489_s_at | 1.11089 | down | 0.006604 | CD44 |
| 204581_at | 1.121758 | down | 0.016833 | CD22 |
| 204666_s_at | 1.105866 | up | 2.69E-04 | RP5-1000E10.4 |
| 205403_at | 1.313192 | up | 0.002585 | IL1R2 |
| 205456_at | 1.104177 | down | 0.0124 | CD3E |
| 206420_at | 1.246161 | up | 0.009528 | IGSF6 |
| 207008_at | 1.262777 | up | 0.010445 | IL8RB |
| 207094_at | 1.125419 | up | 0.023716 | IL8RA |
| 207104_x_at | 1.10934 | up | 0.036019 | LILRB1 |
| 207314_x_at | 1.115032 | down | 0.044444 | KIR3DL2 |
| 207536_s_at | 1.121413 | up | 0.003446 | TNFRSF9 |
| 207643_s_at | 1.168614 | up | 0.009971 | TNFRSF1A |
| 207655_s_at | 1.123825 | down | 0.048559 | BLNK |
| 207697_x_at | 1.153652 | up | 0.00905 | LILRB2 |
| 207827_x_at | 1.351602 | down | 0.019893 | SNCA |
| 207907_at | 1.15163 | up | 2.80E-04 | TNFSF14 |
| 208405_s_at | 1.251242 | down | 0.041759 | CD164 |
| 208485_x_at | 1.117304 | up | 0.0214 | CFLAR |
| 208540_x_at | 1.108202 | up | 0.042068 | S100A11/P |
| 208666_s_at | 1.213137 | down | 1.26E-04 | ST13 |
| 208667_s_at | 1.248621 | down | 3.96E-04 | ST13 |
| 208931_s_at | 1.150311 | down | 0.019467 | ILF3 |
| 209189_at | 1.164217 | up | 0.021235 | FOS |
| 209508_x_at | 1.140008 | up | 0.007207 | CFLAR |
| 209671_x_at | 1.184911 | down | 0.016073 | TRA@ /// TRAC |
| 209723_at | 1.28675 | down | 0.007788 | SERPINB9 |
| 209939_x_at | 1.17078 | up | 0.009053 | CFLAR |
| 210031_at | 1.174872 | down | 0.027308 | CD247 |
| 210042_s_at | 1.265922 | up | 0.048335 | CTSZ |
| 210225_x_at | 1.137269 | up | 0.003548 | LILRB3 |
| 210514_x_at | 1.105079 | up | 0.040499 | HLA-G |
| 210563_x_at | 1.152512 | up | 0.007233 | CFLAR |
| 210564_x_at | 1.141197 | up | 0.011338 | CFLAR |
| 210754_s_at | 1.254778 | up | 0.018888 | LYN |
| 210776_x_at | 1.100981 | down | 0.009593 | TCF3 |
| 210784_x_at | 1.20723 | up | 7.90E-04 | LILRA6 /// LILRB3 |
| 210943_s_at | 1.309776 | up | 0.003924 | LYST |
| 210972_x_at | 1.199937 | down | 0.030357 | TRA@ |
| 211005_at | 1.149716 | down | 0.009554 | LAT /// SPNS1 |
| 211133_x_at | 1.118174 | up | 0.010374 | LILRA6 /// LILRB3 |
| 211135_x_at | 1.184438 | up | 0.001229 | LILRB3 |
| 211163_s_at | 1.210836 | up | 0.005458 | TNFRSF10C |
| 211316_x_at | 1.20165 | up | 0.003518 | CFLAR |
| 211339_s_at | 1.215744 | down | 0.025145 | ITK |
| 211372_s_at | 1.208659 | up | 0.009117 | IL1R2 |
| 211546_x_at | 1.339729 | down | 0.017353 | SNCA |
| 211862_x_at | 1.13062 | up | 0.014222 | CFLAR |
| 211902_x_at | 1.185072 | down | 0.006675 | TRA@ |
| 212414_s_at | 1.224381 | down | 0.018298 | N-PAC /// SEPT6 |
| 213318_s_at | 1.150105 | down | 0.031006 | BAT3 |
| 213894_at | 1.100305 | up | 0.002152 | THSD7A |
| 214359_s_at | 1.19457 | down | 0.037662 | HSP90AB1 |
| 214486_x_at | 1.186758 | up | 0.001419 | CFLAR |
| 214574_x_at | 1.107749 | up | 0.041912 | LST1 |
| 215338_s_at | 1.139026 | up | 0.007906 | NKTR |
| 215967_s_at | 1.108165 | down | 0.012386 | LY9 |
| 216300_x_at | 1.126279 | up | 0.002184 | RARA |
| 217412_at | 1.122067 | down | 0.016503 | TRD@ |
| 217433_at | 1.12529 | up | 0.00102 | TACC1 |
| 217550_at | 1.114039 | up | 0.030685 | ATF6 |
| 217552_x_at | 1.105204 | up | 0.047014 | CR1 |
| 217724_at | 1.163798 | down | 0.039041 | SERBP1 |
| 217775_s_at | 1.149285 | down | 0.003501 | RDH11 |
| 218450_at | 1.16605 | down | 0.043368 | HEBP1 |
| 218831_s_at | 1.131095 | up | 0.022043 | FCGRT |
| 219528_s_at | 1.187628 | down | 0.032716 | BCL11B |
| 219667_s_at | 1.197731 | down | 0.014477 | BANK1 |
| 220088_at | 1.18788 | up | 0.040003 | C5AR1 |
| 220938_s_at | 1.101576 | up | 8.66E-04 | GMEB1 |
| 221478_at | 1.437751 | down | 0.006494 | BNIP3L |
| 221571_at | 1.120485 | down | 0.011251 | TRAF3 |
| 221602_s_at | 1.161923 | down | 0.001686 | FAIM3 |
| 222218_s_at | 1.143145 | up | 0.031853 | PILRA |

References

Alvarez-Erviti, L., Y. Couch, et al. (2011). "Alpha-synuclein release by neurons activates the inflammatory response in a microglial cell line." Neurosci Res **69**(4): 337-342.

Asea, A., S. K. Kraeft, et al. (2000). "HSP70 stimulates cytokine production through a CD14-dependant pathway, demonstrating its dual role as a chaperone and cytokine." Nat Med **6**(4): 435-442.

Bachoud-Levi, A. C., V. Gaura, et al. (2006). "Effect of fetal neural transplants in patients with Huntington's disease 6 years after surgery: a long-term follow-up study." Lancet Neurol **5**(4): 303-309.

Betarbet, R., T. B. Sherer, et al. (2000). "Chronic systemic pesticide exposure reproduces features of Parkinson's disease." Nat Neurosci **3**(12): 1301-1306.

Bian, M. J., L. M. Li, et al. (2009). "Elevated interleukin-1beta induced by 1-methyl-4-phenyl-1,2,3,6-tetrahydropyridine aggravating dopaminergic neurodegeneration in old male mice." Brain Res **1302**: 256-264.

Bjorklund, A., S. B. Dunnett, et al. (2003). "Neural transplantation for the treatment of Parkinson's disease." Lancet Neurol **2**(7): 437-445.

Blum-Degen, D., T. Muller, et al. (1995). "Interleukin-1 beta and interleukin-6 are elevated in the cerebrospinal fluid of Alzheimer's and de novo Parkinson's disease patients." Neurosci Lett **202**(1-2): 17-20.

Brochard, V., B. Combadiere, et al. (2009). "Infiltration of CD4+ lymphocytes into the brain contributes to neurodegeneration in a mouse model of Parkinson disease." J Clin Invest **119**(1): 182-192.

Chen, H., S. M. Zhang, et al. (2003). "Nonsteroidal anti-inflammatory drugs and the risk of Parkinson disease." Arch Neurol **60**(8): 1059-1064.

Chen, S., W. D. Le, et al. (1998). "Experimental destruction of substantia nigra initiated by Parkinson disease immunoglobulins." Arch Neurol **55**(8): 1075-1080.

Dauer, W., N. Kholodilov, et al. (2002). "Resistance of alpha -synuclein null mice to the parkinsonian neurotoxin MPTP." Proc Natl Acad Sci U S A **99**(22): 14524-14529.

Gatto, E. M., M. C. Carreras, et al. (1996). "Neutrophil function, nitric oxide, and blood oxidative stress in Parkinson's disease." Mov Disord **11**(3): 261-267.

Giraud, S. N., C. M. Caron, et al. (2010). "Estradiol inhibits ongoing autoimmune neuroinflammation and NFkappaB-dependent CCL2 expression in reactive astrocytes." Proc Natl Acad Sci U S A **107**(18): 8416-8421.

Griffin, W. S., L. Liu, et al. (2006). "Interleukin-1 mediates Alzheimer and Lewy body

pathologies." J Neuroinflammation **3**: 5.

Hamza, T. H., C. P. Zabetian, et al. (2010). "Common genetic variation in the HLA region is associated with late-onset sporadic Parkinson's disease." Nat Genet **42**(9): 781-785.

He, Y., W. D. Le, et al. (2002). "Role of Fcgamma receptors in nigral cell injury induced by Parkinson disease immunoglobulin injection into mouse substantia nigra." Exp Neurol **176**(2): 322-327.

Huber, V. C., T. Mondal, et al. (2006). "Serum antibodies from Parkinson's disease patients react with neuronal membrane proteins from a mouse dopaminergic cell line and affect its dopamine expression." J Neuroinflammation **3**: 1.

Johnston, L. C., X. Su, et al. (2008). "Human interleukin-10 gene transfer is protective in a rat model of Parkinson's disease." Mol Ther **16**(8): 1392-1399.

Krystkowiak, P., V. Gaura, et al. (2007). "Alloimmunisation to donor antigens and immune rejection following foetal neural grafts to the brain in patients with Huntington's disease." PLoS One **2**(1): e166.

Loeffler, D. A., D. M. Camp, et al. (2006). "Complement activation in the Parkinson's disease substantia nigra: an immunocytochemical study." J Neuroinflammation **3**: 29.

Meredith, G. E., P. K. Sonsalla, et al. (2008). "Animal models of Parkinson's disease progression." Acta Neuropathol **115**(4): 385-398.

Nalls, M. A., V. Plagnol, et al. (2011). "Imputation of sequence variants for identification of genetic risks for Parkinson's disease: a meta-analysis of genome-wide association studies." Lancet **377**(9766): 641-649.

Orr, C. F., D. B. Rowe, et al. (2005). "A possible role for humoral immunity in the pathogenesis of Parkinson's disease." Brain **128**(Pt 11): 2665-2674.

Papachroni, K. K., N. Ninkina, et al. (2007). "Autoantibodies to alpha-synuclein in inherited Parkinson's disease." J Neurochem **101**(3): 749-756.

Pott Godoy, M. C., R. Tarelli, et al. (2008). "Central and systemic IL-1 exacerbates neurodegeneration and motor symptoms in a model of Parkinson's disease." Brain **131**(Pt 7): 1880-1894.

Reynolds, A. D., D. K. Stone, et al. (2010). "Regulatory T cells attenuate Th17 cell-mediated nigrostriatal dopaminergic neurodegeneration in a model of Parkinson's disease." J Immunol **184**(5): 2261-2271.

Ross, G. W., R. D. Abbott, et al. (2000). "Association of coffee and caffeine intake with the risk of Parkinson disease." JAMA **283**(20): 2674-2679.

Saha, R. N. and K. Pahan (2003). "Tumor necrosis factor-alpha at the crossroads of neuronal life and death during HIV-associated dementia." J Neurochem **86**(5): 1057-1071.

Sawada, H. and S. Shimohama (2003). "Estrogens and Parkinson disease: novel approach for neuroprotection." Endocrine **21**(1): 77-79.

Spencer, D. D., R. J. Robbins, et al. (1992). "Unilateral transplantation of human fetal mesencephalic tissue into the caudate nucleus of patients with Parkinson's disease." N Engl J Med **327**(22): 1541-1548.

Vila, M., S. Vukosavic, et al. (2000). "Alpha-synuclein up-regulation in substantia nigra dopaminergic neurons following administration of the parkinsonian toxin MPTP." J Neurochem **74**(2): 721-729.

Wahner, A. D., J. S. Sinsheimer, et al. (2007). "Inflammatory cytokine gene polymorphisms and increased risk of Parkinson disease." Arch Neurol **64**(6): 836-840.

Zhang, W., T. Wang, et al. (2005). "Aggregated alpha-synuclein activates microglia: a process leading to disease progression in Parkinson's disease." FASEB J **19**(6): 533-542.